\documentclass[12pt,preprint]{aastex}
\slugcomment{accepted for publication in ApJ}

\shorttitle{Disk in the X Her}
\shortauthors{J. Nakashima}

\begin{document}

%%%%%%%%%%%
%% title %%
%%%%%%%%%%%

\title{DISK-LIKE STRUCTURE IN THE SEMI-REGULAR PULSATING STAR, X HER}
%\title{KINETIC NATURE OF THE VERY NARROW MOLECULAR LINE TOWARD THE AGB STAR, X Her}
%\title{HIGH-RESOLUTION OBSERVATION OF THE VERY NARROW MOLECULA ROTATIONAL LINE TOWARD SEMIREGULAR PULSATING STAR X HER}

%%%%%%%%%%%%%%%%%%%%%%%%%
%% authors information %%
%%%%%%%%%%%%%%%%%%%%%%%%%

\author{Jun-ichi Nakashima\altaffilmark{1}}

\altaffiltext{1}{Department of Astronomy, University of Illinois at Urbana-Champaign,
1002 W. Green St., Urbana, IL 61801; junichi@astro.uiuc.edu}

%%%%%%%%%%%%%%
%% abstract %%
%%%%%%%%%%%%%%

\begin{abstract}
The author reports a result of an interferometric observation of the semiragular pulsating star with an unusual narrow molecular line profile, X Her, in the CO $J=1$--0 line with the Berkeley-Illinois-Maryland array. In the CO spectrum, a double-component profile (including narrow and broad components) is seen as reported by previous observations. The narrow component consists of two spiky peaks. The spatial structure of the board component shows bipolar shape, and that of the narrow component shows an elliptical/spherical shape. The two peaks in the narrow component show a systematic difference in the integrated intensity map. The kinematical and geometrical properties of the narrow component are reminiscent of a Keplerian rotating disk with the central mass of 0.9 M$_{\odot}$, though an interpretation by an expansion disk seems to be more natural. A secondary bipolar flow instead of the disk cannot be fully excluded as an interpretation of the narrow line. 
\end{abstract}

%%%%%%%%%%%%%%
%% keywords %%
%%%%%%%%%%%%%%

\keywords{stars: AGB and post-AGB ---
(stars:) circumstellar matter ---
stars: imaging ---
stars: individual (X Her) ---
stars: late-type ---
stars: mass loss}

%%%%%%%%%%%%%%%%%%
%% Introduction %%
%%%%%%%%%%%%%%%%%%

\section{INTRODUCTION}
A non-negligible number of asymptotic giant branch (AGB)/post-AGB stars are known to show peculiar molecular line profiles that exhibit unusual small line width. The width of the peculiar narrow line is often less than 5 km s$^{-1}$. Such a narrow line profile is difficult to be understood as an AGB wind with a typical expansion velocity of about 10 km s$^{-1}$. Among CO profiles of AGB stars, the peculiar narrow line profiles are known in 5-10 \% of the sample \citep[see, e.g.,][]{win03}. The narrow line has been found first toward AGB stars with double-component line profiles \citep{kna98,ker99}; in these envelopes, the narrow and broad lines have the same central velocity. The narrow lines are so far found in several types of AGB stars. For instance, \citet{kna98} have reported the narrow lines toward both M- and C-type AGB stars. \citet{ker99} also have found the narrow lines toward four semi-regular pulsating AGB stars. \citet{kah96} have found similar narrow lines toward chemically peculiar AGB stars, BM Gem and EU And; these stars are known as "{\it silicate carbon stars}", which simultaneously have a C-rich central star and O-rich circumstellar material. \citet{nak04a} have recently reported a detection of the narrow line toward an unusual SiO maser source with a rich set of molecular species, IRAS 19312+1950, and they have suggested this object is also an AGB/post-AGB star, though the evolutional status of this object is not definitely known yet. In addition, AGB stars with only the narrow component (without the broad component) are also known \citep{ker99,win03}, but these are chemically normal. To explain the origin of the double-component profile including the narrow line, \citet{kna98} have advanced a hypothesis taking account of multiple shell winds, in which each shell has different expanding velocities produced by episodic mass loss with highly varying gas expansion velocities. In addition, \citet{kah96} have suggested that the narrow lines seen in the silicate carbon stars are explained by gravitationally stable material in a form of a distorted or puffed-up slowly rotating disk, in which O-rich material is trapped. Interestingly, \citet{ber00} have reported a tentative detection of a Keplerian disk toward an O-rich AGB star with the narrow line, RV Boo. 

To investigate the properties of peculiar narrow lines in AGB stars, high-resolution interferometric observations were made with the Berkeley-Illinois-Maryland Association (BIMA) array toward a sample of AGB stars exhibiting the narrow molecular line profiles. In this paper, the author reports the first result of the interferometric observations in the CO $J=1$--0 line toward the semiragular pulsating star with normal O-rich chemistry, X Her, and shows an indication of disk structure around the central star. X Her (IRC+50248; SAO 45863; IRAS 16011+4722) is a semi-regular pulsating AGB star having the typical narrow line; this star has been reported in the {\it General Catalog of Variable Stars} \citep{kho85} to have a period of 95 days. There exists a second period of 746 days \citep{hou63}. The double-component profile of X Her has been found first by \cite{kah96}, and they also have provided the first spatial information of X Her through their single-dish mapping observation. \citet{kah96} have found that the broad symmetric line of X Her consists of a bipolar flow, and that the overall spatial structure of the narrow line is almost spherical. The annual parallax of this star (7.26 mas with 1$\sigma$ uncertainty of 0.70 mas) was measured by the Hipparcos satellite, giving the distance of 126--152 pc. In this paper the distance of 138 pc is applied.

%%%%%%%%%%%%%%%%%
%% Observation %%
%%%%%%%%%%%%%%%%%

\section{OBSERVATION AND RESULTS}
Interferometric observations of X Her were made with the BIMA array from 2004 February to May. The instrument was described in detail by \citet{wel96}. The author observed the CO $J=1$--0 line at 115.271202 GHz with the BIMA array consisting of 10 elements in 4 configurations (A, B, C and D). The observations were interleaved every 25 (for B, C and D arrays) or 10 minutes (for A-array) with the nearby point sources, 3C345 and 1613+342, to track the phase variations over time. The absolute flux calibration was determined from observations of Uranus and MWC349, and is accurate to within 20\%. The final map has an accumulated on-source observing time of about 31 hr. Typical single-sideband system temperature ranges from 250 to 500 K. The velocity coverage is about 350 km s$^{-1}$, using three different correlator windows with a band width of 50 MHz each. The velocity resolution is 1.03 km s$^{-1}$. The phase center of the map is R.A.$=$16$^h$02$^m$39.1739$^s$ , decl.$=$47$^{\circ}$14$'$25.279$''$ (J2000). Data reduction was performed with the MIRIAD software package \citep{sau95}. Standard data reduction, calibration, imaging, and deconvolution procedures were followed. Robust weighting of the visibility data gives a  $3.11'' \times 2.48''$ CLEAN beam with a position angle of 37.18$^{\circ}$. The rms noise per 1.0 km s$^{-1}$ is 3.71$\times$10$^{-2}$ Jy beam$^{-1}$.

Figure 1 shows the spatially integrated spectrum of the CO $J=1$--0 line. In the spectrum two kinematical components are seen as reported by \citet{kah96}: "broad wing component" ($V_{lsr}$ range: $-82$ -- $-63$ km s$^{-1}$) and "narrow component" ($V_{lsr}$ range: $-76$ -- $-70$ km s$^{-1}$) Furthermore, the narrow component consists of two kinematical components: "blue-shifted peak" ($V_{peak} \sim -75$ km s$^{-1}$) and "red-shifted peak" ($V_{peak} \sim -72$ km s$^{-1}$). These kinematical components are indicated by the arrows in Figure 1. The continuum radiation is not detected at the 3$\sigma$ rms level, though the frequency channels were integrated over 150 MHz using the opposite-side band to the line observation of CO. The upper limit of the flux of the continuum radiation at the 3 mm band (center frequency is $\sim$112 GHz) is $6.2 \times 10^{-3}$ Jy. To investigate the spatial distribution of each kinematical component, velocity integrated intensity maps were made; the maps are shown in Figure 2. The structure seen in Figure 2 is clearly resolved by our synthesized beam. According to the peak intensity of the CO spectrum (0.18 Jy beam$^{-1}$), about 5\% of the flux detected by the IRAM 30-m telescope \citep{kah96}, was detected in the present interferometric observation. As reported by \citet{kah96}, in the upper panel of Figure 2 we can clearly see the bipolar shape consisting of the blue- and red shifted wings of the broad component (thin and thick contours), and also can see the spherical shape consisting of the narrow component (gray broken contours). The angular size of the bipolar shape (above 5$\sigma$ level) is 12.4$''$ (length) and 11.8$''$ (width); this angular size corresponds to the linear size of $2.6 \times 10^{16}$ cm (length) and $2.4 \times 10^{16}$ cm (width) at the distance of 138 pc. Strictly speaking, according to our high-resolution map, the spatial structure of the narrow component seems to be elliptical rather than spherical (this is more remarkable in the lower panel of Figure 2 [see the next paragraph]). The best result of fitting by elliptic curves to the 5 $\sigma$ level contour, which is expected to be undisturbed by an effect of the beam pattern, gives the angular sizes of major and minor axes of the elliptical shape, 15.1$''$ and 12.6$''$, respectively; these angular sizes correspond to the linear sizes of $3.1 \times 10^{16}$ cm (major axis) and $2.6 \times 10^{16}$ cm (minor axis) at the distance of 138 pc. 

The lower panel of Figure 2 shows velocity integrated intensity maps of the blue- and red-shifted peaks. Interestingly, in the lower panel of Figure 2 we can clearly see a systematic difference between the shapes of the thin and thick contours. This phenomenon is also confirmed in velocity channel maps; the maps are shown in Figure 3. In Figure 3 the top, middle and bottom panels correspond to the blue-shifted wing, the narrow component and the red-shifted wing, respectively. The channel maps clearly show that the CO source structure varies with velocity. The velocity structure of the narrow component seen in Figure 3 (middle panels) seems to be varying systematically, while the variation of the structure at the higher-velocity ranges seen in the upper ($-82$ -- $-80$ km s$^{-1}$) and bottom ($-68$ -- $-66$ km s$^{-1}$) panels seems to originate in the bipolar flow reported by \citet{kah96}.

%%%%%%%%%%%%%%%%
%% Discussion %%
%%%%%%%%%%%%%%%%

\section{DISCUSSION}
To precisely investigate the kinematical properties of the narrow component, position--velocity ($p$-$v$) diagrams were made in various cuts with various position angles. Figure 4 shows two selected $p$-$v$ diagrams. The origin of the cuts are taken at the apparent symmetric center of the spatial structure of the narrow component: (RA, decl.,)$_{\rm offset}$ = (1.0$''$, 0.1$''$). The cuts used in Figure 4 are indicated by the dotted arrows in Figure 2. The most remarkable velocity structure has been found in the position angle of 61$^{\circ}$ (cut A), and on the other hand, the velocity structure almost totally disappears in the position angle of 151$^{\circ}$ (cut B), which is perpendicular to the cut A. In the upper panel of Figure 4, we can clearly see a systematic variation of structure as a function of the radial velocity. Although the velocity structure originating in the bipolar flow seems to contaminate to the data especially in higher-velocity ranges, the structure of the narrow component is symmetrically placed in velocity with respect to the line center ($V_{sys} \sim 73.5$ km s$^{-1}$). This kinematical structure is reminiscent of Keplerian rotation \citep[see, e.g., Figure 8 in][]{oha97}. The structure can be fitted well actually by a Keplerian rotation curve with the central mass of 0.9 M$_{\odot}$ at the distance of 138 pc; this Keplerian rotation curve is superimposed on the upper panel of Figure 4. The central mass calculated by the Keplerian fitting is to the point as a mass of the star of this type if we rely on edge on view. If the elliptical shape seen in Figure 2 represents the true geometry of the disk, we can crudely estimate the orbital-inclination angle of the Keplerian disk on the assumption of thin-disk geometry. In such a case, with the length of the major and minor axes gave in the previous section, the orbital-inclination angle of the Keplerian disk is calculated to be 57$^{\circ}$; this angle leads to the corrected central mass, 3.0 M$_{\odot}$. This corrected mass seems to be somewhat heavier than that expected for X Her, because the luminosity of AGB stars, in general, varies as a function of a main-sequence mass of the star, and also because the luminosity of X Her is typical as an AGB star with a main-sequence mass of about 1 M$_{\odot}$ \citep{vas93}. In fact, the total luminosity of X Her is estimated to be about 9200 L$_{\odot}$ from V and I$_{\rm c}$-band fluxes \citep{pla03}, J, H, K, L' and M-band fluxes \citep{ker94} and the IRAS 12 and 25 $\mu$m fluxes on the basis of a linear interpolation \citep{deg02}. Therefore, if the narrow component really originates in a Keplerian rotating disk, in my opinion, an interpretation in the form of a geometrically thick disk, which is observed in closely edge-on view, is tempting. One problem remaining in the Keplerian disk interpretation is the physical/geometrical relationship between the bipolar flow and the disk. If the axes of the bipolar flow and the Keplerian disk correspond, the kinematical properties seen in the present results are rather difficult to be understood unless we assume existence of outflow (or expansion) in the disk. In fact, a simple expanding disk model suggested by \citet{hir04} seems to, on some level, explain the integrated intensity maps shown in Figure 2 \citep[compare Figure 2 in this paper and Figure 4 in][]{hir04}, although the velocity scale of their model is totally different from the present case. The geometrical correspondence of axes between bipolar flow and disks is frequently required in recent AGB/post-AGB studies to explain the formation of the bipolar flows seen in AGB/post-AGB stars. If the expansion disk is the case, there remains no indication of Keplerian rotation.

A secondary bipolar flow instead of expansion of the disk might be an alternative to explain the narrow component. However, in such a case, the very slow-expansion velocity ($\sim$1.5 km s$^{-1}$) is problematic unless it is a projection effect, because recent studies suggest that the bipolar flow of AGB/post-AGB stars originates in an AGB wind, and also because, in such a case, the velocity of the bipolar flow is expected to be at least typical of an AGB wind (10--20 km s$^{-1}$) or higher than that. If the slow expansion velocity originates in the projection effect, the inclination angle of the secondary bipolar flow is estimated to be about 81$^{\circ}$ on the assumption that the expansion velocity of the secondary bipolar flow is equivalent to a half of the full-line width of the broad component ($\sim 10$ km s$^{-1}$). In such a large inclination, the elliptical shape that is almost spherical is unlikely to be explained by the bipolar flow if the true structure of the bipolar flow is well collimated. However, we have to note that, in terms of the dynamical age, the secondary bipolar flow may be produced within the duration of the AGB phase in the evolution of X Her, because the dynamical time scale of the secondary bipolar flow is estimated to be 6500 year; the order of this value is equivalent to the value calculated by \citet{kah96}. Thus, at the moment, the possibility of the secondary bipolar flow cannot be fully excluded. We should note that \citet{ker03} have recently presented an SiO (v=0, J=1--0) map taken by the Very Large Array (VLA); the map also suggests a Keplerian rotating disk around X Her, but in a direction orthogonal to the disk in this paper, though the data of \cite{ker03} are marginal quality. To draw a more firm conclusion on this matter, higher-resolution and higher-sensitivity data are required.

Although several AGB stars with narrow lines exhibit chemical peculiarity, no evidence of such peculiarity has been so far found in X Her. \citet{gon03} have detected both the broad and narrow kinematical components toward X Her in the thermal millimeter line of the SiO molecule. They have estimated the column density of SiO, but have found no peculiarity in SiO abundance; it is fairly normal as an O-rich AGB star. The IRAS LRS class of X Her is 24; this also suggests X Her is a normal O-rich star with emission feature of silicate. As a possible interpretation of the normality of the chemical condition of X Her, the author suggests that the kinematical properties of the (possible) disk is critical for the chemistry of the star. The key is the duration of material in the disk; this duration is constrained by the kinematical properties of the disk. Thus, if the expansion is dominant in the disk, the disk cannot effectively trap material for a long time; adversely, if the Keplerian rotation is dominant, the disk may effectively trap plenty of material for a relatively long time. Therefore, if the chemical environment of the star changes during the duration, the star may have complex (or multiple) chemical environments, such as the C/O-rich mixture seen in silicate-carbon stars. This mechanism has been first discussed by \citet{jur99}. According the present results, the disk of X Her seems to include expanding motion (though other possibilities cannot be excluded). Therefore, the material included in the disk of X Her is limited to that expelled recently. To confirm this hypothesis, it is hoped that AGB/post-AGB stars with the narrow line and chemical peculiarities, such as silicate carbon stars, is observed by high-resolution radio interferometry, and that the results is compared with those of chemically normal AGB stars with the narrow line, such as X Her. Generally speaking, intensities of the narrow lines seen in AGB stars are very weak, and is difficult to be observed with current equipments, but next generation interferometers will enable us to increase the number of observations of AGB stars with narrow lines.

%%%%%%%%%%%%%
%% Summary %%
%%%%%%%%%%%%%

\section{SUMMARY}
In this paper, the author reports a result of an interferometric observation of the semiragular pulsating AGB star, X Her, in the CO $J=1$--0 line. In the CO spectrum, a double-component profile, including narrow and broad components, is seen as reported by \citet{kah96}. The narrow component consists of two peaks; these peaks shows a systematic difference in the integrated intensity map. The kinematical properties of the narrow component is reminiscent of a Keplerian disk with the central mass of 0.9 M$_{\odot}$, though the possibilities of a bipolar flow and expanding disk instead of the Keplerian motion cannot be excluded.

%%%%%%%%%%%%%%%%%%%%%
%% Acknowledgments %%
%%%%%%%%%%%%%%%%%%%%%

\acknowledgments
This research has been supported by the Laboratory for Astronomical Imaging at the University of Illinois and by NSF grant AST 0228953, and has made use of the SIMBAD and ADS databases.

\begin{figure}
\epsscale{.60}
\plotone{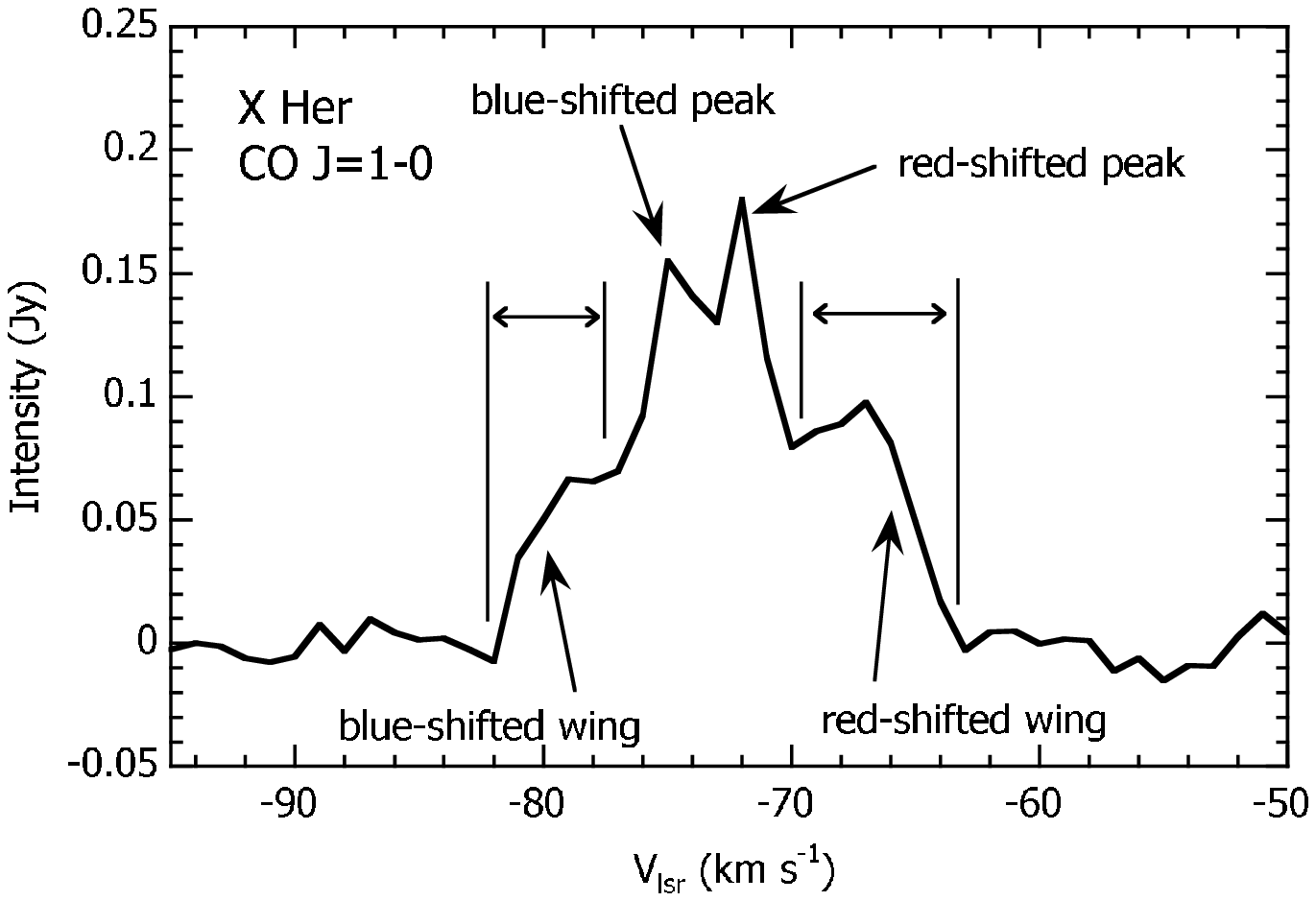}
\caption{Spatially integrated spectrum of X Her in the CO $J=1$--0 line. The integrated area is a circle with a diameter of 15$''$. The vertical solid lines represent the velocity ranges of the blue- and red-shifted wings. Each kinematical component is indicated by the arrows. \label{fig1}}
\end{figure}

\clearpage

\begin{figure}
\epsscale{.40}
\plotone{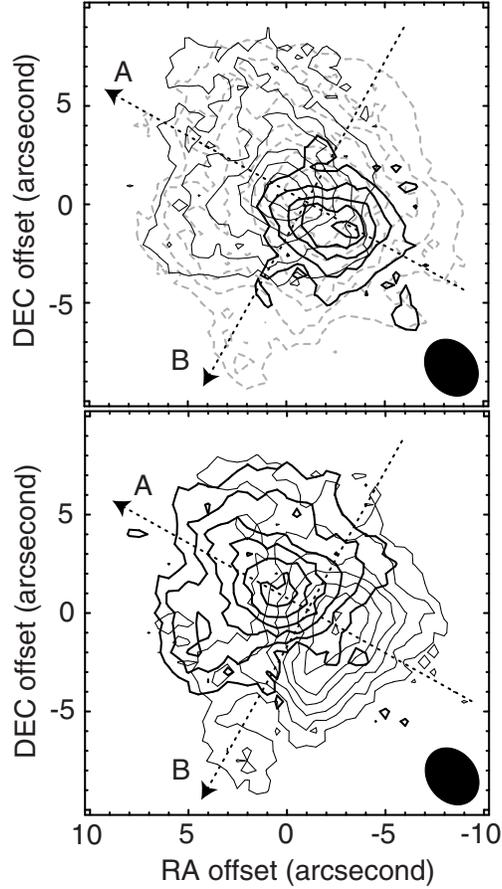}
\caption{Velocity integrated intensity maps of the kinematical components indicated in Figure 1. In the upper panel, the thick and thin contours represent the maps of the blue- and red-shifted wings, respectively; the gray broken contours represent the map of combined intensity of the blue- and red-shifted peaks. In the lower panel, the thick and thin contours represent the maps of the blue- and red-shifted peaks, respectively. The dotted arrows represent the cuts used for Figure 3. The synthesized beam size is indicated in the lower-right corners. The contours start from a 5$\sigma$ level, and the increment of the contours are every 2$\sigma$. In the upper panel the 1 $\sigma$ levels for the thick, thin and broken contours are $1.516 \times 10^{-2}$, $1.238 \times 10^{-2}$ and $1.403 \times 10^{-2}$ Jy beam$^{-1}$, respectively. Similarly, in the lower panel, the 1$\sigma$ level is $2.144 \times 10^{-2}$ Jy beam$^{-1}$ for both the thick and thin contours. The velocity integration ranges for the wings are $-82$ --- $-77$(blue-shifted wing) and $-71$ --- $-63$(red-shifted wing), and those of the red- and blue-shifted peaks are 3 km s$^{-1}$ each (the peak velocities are taken at the center of the range). \label{fig2}}
\end{figure}

\clearpage

\begin{figure}
\epsscale{.70}
\plotone{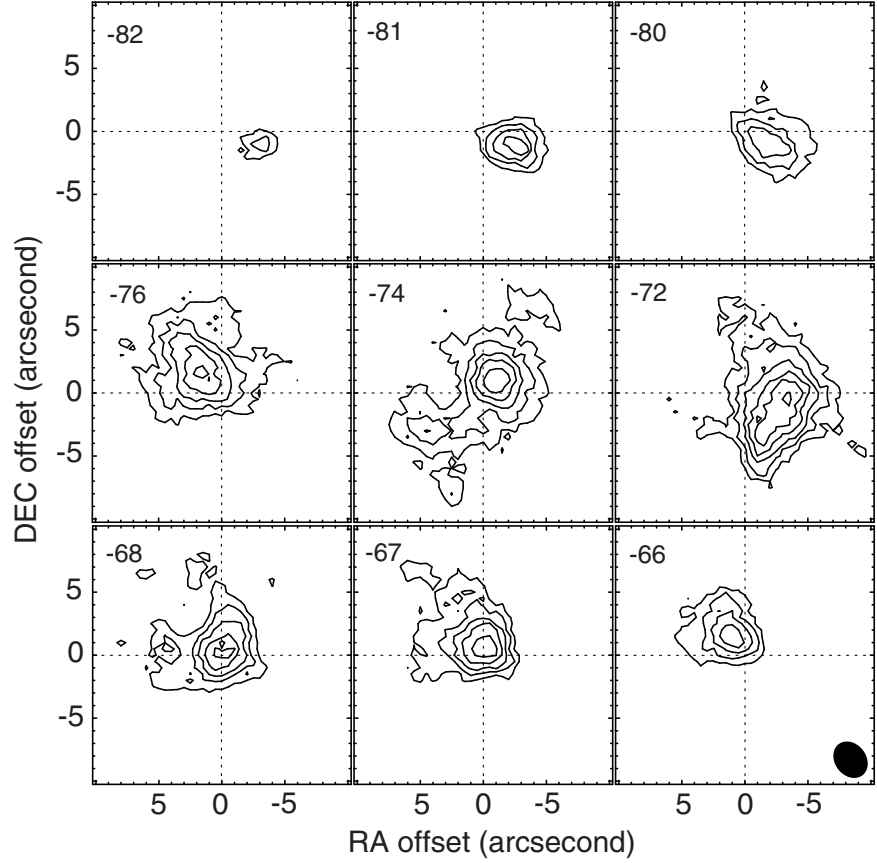}
\caption{Selected channel maps of the CO $J=1$--0 line. The channel velocities are indicated in the upper left corner of each panel. The velocity channels were averaged over 2 km s$^{-1}$ intervals. The contours start from 5$\sigma$ level, and the increment of the contours is every 2$\sigma$. The 1$\sigma$ level corresponds to $2.63 \times 10^{-2}$ Jy beam$^{-1}$. The synthesized beam size is indicated in the lower-right corner. \label{fig3}}
\end{figure}

\clearpage

\begin{figure}
\epsscale{.50}
\plotone{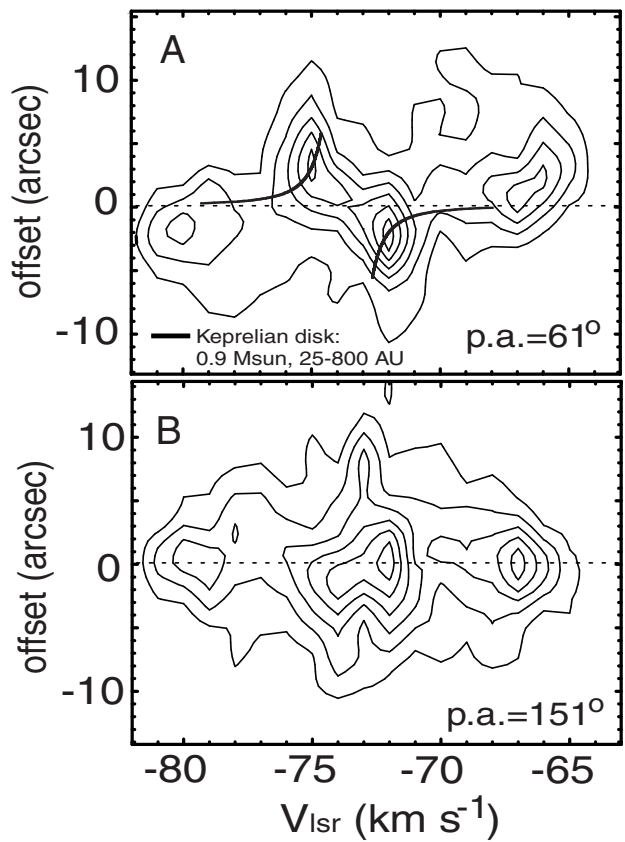}
\caption{Position--velocity diagrams for the cuts indicated in Figure 2. The contour levels correspond to 90, 75, 60, 45, 30 and 15\% of the intensity peak. The names of the cuts and the position angles are respectively indicated in the upper-left and lower-right corners of each panel. The solid curve indicated in the upper panel represents the best fit result of the Keplerian disk to the data (see text). The broken horizontal lines represent the origin of the offset axes. \label{fig4}}
\end{figure}

%%%%%%%%%%%%%%%%%%%%%%%%%%%%%%
\end{document}